\documentclass[12pt,a4paper]{article}
\usepackage[left=1cm,right=1cm,top=2cm,bottom=2cm,nohead]{geometry}
\usepackage{amsmath,amssymb}
\numberwithin{equation}{section}
\usepackage{hyperref}
\usepackage{graphicx}
\usepackage{array}
\usepackage{booktabs}
\usepackage{longtable}
\usepackage{enumitem}
\usepackage{color,soul}
\usepackage[utf8]{inputenc}
\usepackage{makecell}
\usepackage{multirow}
\usepackage{authblk}
\usepackage{lscape}
\usepackage[raggedrightboxes]{ragged2e}
\paperwidth=\dimexpr \paperwidth + 7cm\relax
\oddsidemargin=\dimexpr\oddsidemargin + 3cm\relax
\evensidemargin=\dimexpr\evensidemargin + 3cm\relax
\marginparwidth=\dimexpr \marginparwidth + 3cm\relax

\usepackage[dvipsnames]{xcolor}

\usepackage[font=small,labelfont=bf]{caption}

\usepackage{apacite}

\usepackage{xr} 
\makeatletter
\newcommand*{\addFileDependency}[1]{
\typeout{(#1)}
\@addtofilelist{#1}
\IfFileExists{#1}{}{\typeout{No file #1.}}
}\makeatother

\usepackage[colorinlistoftodos, textwidth=36mm, textsize=singlespacetiny, shadow]{todonotes}

\usepackage{setspace}
\def\appendix{\textit{Supporting Information}}

\begin{document}

\title{Using language models to label clusters of scientific documents}

\author[a]{Dakota Murray}
\author[b]{Chaoqun Ni}
\author[b]{Weiye Gu}
\author[c]{Trevor Hubbard}
\affil[a]{Department of Information Science \& Technology, University at Albany, State University of New York, Albany, NY, USA}
\affil[b]{Information School, University of Wisconsin - Madison, Madison, WI, USA}
\affil[c]{Digital Science, Cambridge, MA, USA}


\maketitle

\clearpage
\begin{abstract}
Automated label generation for clusters of scientific documents is a common task in bibliometric workflows. 
Traditionally, labels were formed by concatenating distinguishing characteristics of a cluster's documents; while straightforward, this approach often produces labels that are terse and difficult to interpret. 
The advent and widespread accessibility of generative language models, such as ChatGPT, make it possible to automatically generate descriptive and human-readable labels that closely resemble those assigned by human annotators. 
Language-model label generation has already seen widespread use in bibliographic databases and analytical workflows. 
However, its rapid adoption has outpaced the theoretical, practical, and empirical foundations.
In this study, we address the automated label generation task and make four key contributions:
(1) we define two distinct types of labels: \textit{characteristic} and \textit{descriptive}, and contrast descriptive labeling with related tasks;
(2) we provide a formal description of the descriptive labeling task that clarifies important steps and design considerations;
(3) we propose a structured workflow for label generation and outline practical considerations for its use in bibliometric workflows; and
(4) we develop an evaluative framework to assess descriptive labels generated by language models and demonstrate that they perform at or near characteristic labels, and highlight design considerations for their use. 
Together, these contributions clarify the descriptive label generation task, establish an empirical basis for the use of language models, and provide a framework to guide future design and evaluation efforts.
\end{abstract}

\newpage
\section*{Introduction}
A common step in bibliometric analysis involves automatically identifying clusters of topically related scientific documents and assigning each a meaningful and human-readable label. These labels help users---who may not be domain experts---to understand the topical focus of the cluster. Label generation, however, is not straightforward.
Traditional approaches often concatenate distinguishing characteristics of documents in a cluster to form a label;
in the case of scientific documents, these are often noun phrases in paper titles, paper abstracts, or the names of publication venues.
For example, a cluster of documents on a specific branch of experimental particle physics might be labeled ``\textit{Standard Model; Higgs Boson; Subatomic Particles; Particle Collider; LHC}''. 
In this paper, we refer to this as a \textbf{\textit{characteristic label}}. Although characteristic labels are relatively easy to generate automatically from the content of the clustered documents, they  tend to be terse and laden with technical jargon. In other words, characteristic labels do not resemble the kind of descriptive summaries that a human annotator would produce.

For most labeling tasks, it is preferable to generate \textbf{\textit{descriptive labels}}: labels that describe and summarize the thematic contents of a document cluster without necessarily including any characteristic terms drawn directly from the documents.
In other words, descriptive labels resemble the types of intuitive and interpretable labels that human annotators would produce.
For example, descriptive labels for clusters of astrophysics papers might include  ``Particle Physics'', ``Galactic Dynamics'', and ``Black hole physics''.
Traditionally, descriptive labels are generated by human experts, making it a time-consuming, intensive, and cognitively demanding process. This dependence on expert input is impractical for large-scale bibliometric applications, where thousands of clusters may require labeling. 

The advent of large generative large language models (LLMs), such as ChatGPT, have made it possible to automate the production of descriptive labels. 
These models have already demonstrated utility in a variety of tasks relevant to social science research~\cite{ziems2024llm-css}, including text summarization~\cite{zhang2024summarization, liu2023summarization, pu2023summarization, liu2024learning} and text annotation~\cite{gilardi2023annotation, kuzman2023annotation, huang2023annotation, tornberg2023annotation}.
Label generation bears a resemblance to these tasks, and so it stands to reason that language models may prove equally useful for automatically creating descriptive labels for arbitrary clusters of scientific documents.
Indeed, with a basic prompt and summary information about a cluster’s contents, a language model can generate descriptive labels that closely resemble those produced by human experts. 
At first glance, such labels are more interpretable and user-friendly than characteristic labels, making them especially valuable for conveying the thematic focus of document clusters.

Descriptive labeling with language models has already been widely adopted by the bibliometric community.
At the time of writing, bibliographic databases such as \textit{OpenAlex}~\footnote{\url{https://help.openalex.org/hc/en-us/articles/24736129405719-Topics}} and \textit{CiteSpace}~\footnote{\url{https://citespace.podia.com/blog/49def10d-9057-4cae-9deb-e75dcee5c9ae}} as well as bibliometric analyses~\cite{eck2024classifying} have incorporated language models to generate labels for clusters of granular topics.
Language models satisfy a need in bibliometric workflows and their adoption is likely to continue. 

Although general-purpose language models show clear potential for automatic label generation, their adoption has outpaced their theoretical, practical, and empirical foundations.
In terms of theory, descriptive labeling lacks a clear definition and distinction from characteristic labeling, and the automated descriptive labeling task is not well specified.
In terms of practical guidance, each LLM-based labeling solution is bespoke, with
researchers and organizations reliant on informal craft knowledge that is neither formalized nor communicated.
Empirically, although LLM-generated descriptive labels are \textit{prima facie} preferable to characteristic labels, there is little quantitative evidence contrasting their quality.
In sum, these gaps highlight the need for a more rigorous conceptual, methodological, and empirical foundation to support the responsible and effective use of LLMs for automated descriptive label generation in bibliometric applications.
Our paper aims to provide this foundation. 

We also aim to address two empirical questions about the use of LLM-based labeling:
(1) What options are there for the design of LLM label generation systems, and to what extent do choices affect the labels produced; and (2) Are descriptive labels comparable to characteristic labels in their ability to distinguish clusters?

This study offers four key contributions.
First, we clarify the distinction between the two types of labels: 
\textit{characteristic labels} represent labels generated through traditional techniques, whereas \textit{descriptive labels} are those of the form produced by language models. 
This clarification of terminology highlights the unique utility of language models for label generation and the challenges inherent in directly comparing them to conventional methods.
We also distinguish descriptive labeling from related tasks such as text summarization, annotation, and classification, highlighting important points of differentiation. 
Second, we provide a formalized description of the descriptive labeling task, which clearly details the various components of labeling workflows and points at which design choices can be made.
Third, we outline a practical workflow for using LLMs to automatically generate labels for clusters of documents, making use of our formalization of the task to identify key decisions to consider when implementing custom workflow.
Fourth, we develop a framework for evaluating the quality of descriptive labels.
We apply this framework to case studies across diverse disciplinary contexts,
seeking to answer the empirical questions detailed previously.
To do so, we develop a series of metrics to measure the impact of prompt design decisions on output labels and human annotators to compare the quality of descriptive labels against their characteristic label counterparts.
Collectively, these contributions establish a robust foundation for applying language models to descriptive label generation and provide a pathway for future methodological improvements in bibliometric analysis.

\section*{Cluster Labeling}
\subsection*{Characteristic and descriptive labeling}
We introduce a distinction between what we term \textit{characteristic} and \textit{descriptive} labels for clusters of documents.
A \textit{characteristic label} is constructed by extracting distinguishing characteristics from documents within a cluster, typically concatenating these characteristics into a short string that forms the final label.
This often involves identifying important terms from the titles and abstracts of documents.
For each cluster, a subset of terms is selected to form the label based on their 'scores'. 
Scoring methods vary, but generally depend on the frequency of terms within the cluster and the distinctiveness relative to other clusters;
terms that are highly frequent in one cluster but rare elsewhere tend to receive the highest scores~\cite{koopman2017mutual, sjgrde2021labeling, chen2010cocitation,Mei2007}. 
Some approaches also incorporate semantic similarity between terms~\cite{li2015labeling}.
Other distinctive features can also serve as the basis for characteristic labels. 
For example, frequent or representative text~\cite{marchetti2020topic}, distinctive journal titles~\cite{velden2017mapping,lamers2021disagreement}, or author-provided keywords~\cite{vahidnia2020mapping} are common.
More holistic approaches, such as the ``CSET Map of Science'' (as of December 2024), include a combination of key terms, high-level subject categories, and other metadata that collectively help users make sense of each cluster~\footnote{\url{https://cset.georgetown.edu/publication/cset-map-of-science/}}. 
Despite methodological variations, the core procedure is clear: the distinctive features of the documents in a cluster are automatically extracted, scored, and the highest scoring features are formed into a label.

Whereas characteristic labels can be precisely defined by an algorithm, descriptive labels lack such an explicit definition.
Here, we adopt an informal definition of descriptive labels as a label that resembles what a human expert might produce when tasked with labeling a cluster. 
Descriptive labels should be summative, adequately describing, to the greatest extent possible, the contents of a cluster.
Descriptive labels should also be intelligible and concise, such that they can be quickly read and understood even by non-expert users.
Descriptive labels should be interpretable, identifying the topic of a cluster within an implicit disciplinary taxonomy; for example, upon encountering the descriptive label ``Solar system dynamics'', even non-expert users should surmise that the cluster represents a sub-field of \textit{Astrophysics} and \textit{Physics}.

The distinction between characteristic and descriptive labels is best illustrated by way of example.
Table~\ref{table:descritve-vs-characteristic} presents characteristic labels for clusters of Astrophysics documents from Koopman and Wang~\cite{koopman2017mutual}, alongside descriptive labels we generated using ChatGPT using the approach outlined later in this paper.
Characteristic labels tend to consist of strings of technical terminology.
Even when the terms are familiar to the reader, they may not clearly convey a coherent topic of study. 
In contrast, descriptive labels use far less technical jargon, offering an accessible summary of the topic that can be understood by an informed reader without deep familiarity with the specialized content of the field.
Although descriptive labels are more readable and interpretable, characteristic labels offer greater specificity. 
By prioritizing distinctive terms through scoring methods, characteristic labels help mitigate issues such as ambiguity or the presence of vague or imprecise terminology.
Moreover, whereas characteristic labels are derived directly from key terms of document characteristics, a descriptive label may not contain any terms from the clustered documents. Instead, a descriptive label can include external terms, but nevertheless effectively capture the overall content or theme of the documents.
For example, cluster 5 in Table~\ref{table:descritve-vs-characteristic} has the descriptive label ``Particle Physics'' and yet the term ``particle'' does not appear among the most distinctive terms of its documents.

\begin{figure}[h!]
    \centering
    \includegraphics[width=1.0\linewidth]{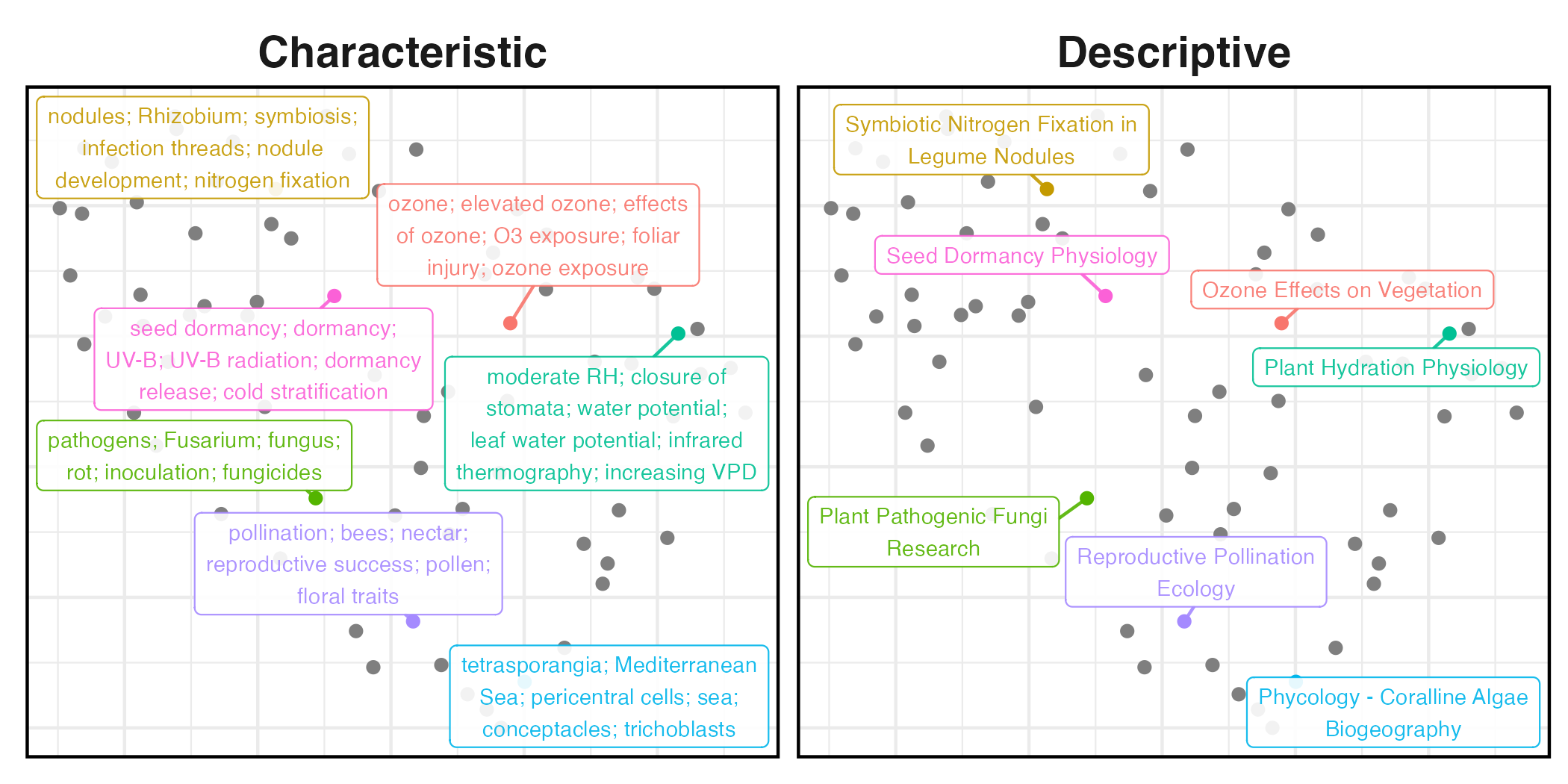}
    \caption{
    \textbf{Example of ``map of science'' comparing cluster labels.}
    Shown is a 2-dimension projection of an embedding demonstrating the practical difference between characteristic (left) and descriptive labels (right). 
    Clusters are identified for the ``Botany'' cluster described in the Methods. 
    We first create embeddings based on publication text (title and abstract) which are passed to \textit{Lingo4G}---a software product aimed at generating maps of documents---and the embeddings are reduced to 2 dimensions for visualization using UMAP~\protect\cite{mcinnes2018umap}.
    }
    \label{fig:umap-example}
\end{figure}

We further illustrate the practical difference between these two types of labels in Fig.~\ref{fig:umap-example} for a set of documents relating to Plant biology, which shows their use in a common ``map of science'' style visualization.
Here, it is apparent that descriptive labels have a clear legibility advantage over their characteristic counterparts; they are concise and quickly communicate the contents of a cluster. 

%
%
\begingroup
\renewcommand{\arraystretch}{1.75}
\setlength{\tabcolsep}{12pt}
\begin{table}[p]
\small
\caption{
Examples of characteristic cluster labels derived from the frequent and distinctive terms of Astrophysics papers. 
Characteristic labels sourced from Koopman \& Wang (2017). 
Descriptive labels are generated using ChatGPT 3.5, using a simple prompt that describes the labeling task and provides the characteristic label.
}
\label{table:descritve-vs-characteristic}
\resizebox{\textwidth}{!}{%
\begin{tabular}{p{0.45in}|p{4in}|p{1.5in}}
\hline
\textbf{Cluster} & \textbf{Characteristic labels} & \textbf{Descriptive labels} \\ \hline
\centering 1 & Solar wind, magnetosphere, interplanetary magnetic, magnetic field, auroral, plasma, magnetopause, ion, substorm, spacecraft &
  Magnetospheric physics \\
\centering  2 & Lens, microlensing, gravitational lens, rotation curve, spiral galaxies, bars, dark matter, barred galaxies, galaxy, pattern speed &
  Galactic Dynamics \\
\centering  3 & Seyfert 1, active galactic, narrow line, agn, broad line, galactic nuclei, quasars, line seyfert, nuclei agns, emission line &
  Active Galactic Nuclei \\
\centering  4 & Spacetimes, black hole, horizon, asymptotically flat, reissner nordstrom, metric, einstein maxwell, spherically symmetric, hole solutions, schwarzschild &
  Black hole physics \\
\centering  5 & Standard model, higgs, lhc, minimal supersymmetric, supersymmetric standard, neutrino mass, lepton, right handed, hadron collider, electroweak &
  Particle physics \\
\centering  6 & Mars, titan, ice, water, deposits, cassini, co2, methane, atmosphere, surface &
  Planetary Science \\
\centering  7 & Galaxy clusters, dark matter, haloes, cluster, n body, weak lensing, intracluster medium, halo mass, 1 mpc, galaxies &
  Cluster Cosmology \\
\centering  8 & Blazar, bl lac, jet, radio sources, lac objects, radio galaxies, synchrotron, radio, flat spectrum, 3c &
  Extragalactic Radio Astronomy \\
\centering  9 & Performance, scientific, technology, mission, astronomical, development, research, flight, cost, software &
  Space Mission Engineering \\
\centering 10 & Globular clusters, fe h, metal poor, red giant, metallicity, giant branch, horizontal branch, galactic globular, color magnitude, stars &
  Stellar Populations \\
\centering  11 & Asteroid, comet, body problem, orbits, kuiper belt, main belt, bodies, mean motion, planets, solar system &
  Solar system dynamics \\ \hline
\end{tabular}%
}
\end{table}
\endgroup

Although there is extensive research on automated cluster labeling methods~\cite{koopman2017mutual,sjgrde2021labeling,chen2010cocitation,velden2017mapping,suominen2015map} and interpretable topic modeling~\cite{marchetti2020topic,role2014labeling,chang2009topics}, prior work has not systematically distinguished between characteristic labels that concatenate cluster characteristics versus those that provide human-interpretable category descriptions.
This distinction has remained implicit in the literature. 
Traditional automated labeling methods typically evaluate their ``characteristic'' labels against human-curated ground truth labels without explicit recognition that these represent fundamentally different approaches to cluster labeling. 
Recent studies have begun to use large language models to generate human-like labels~\cite{kozlowski2024generative,khandelwal2025}, yet they do not explicitly theorize the distinction between label types, nor the implications of this new label type for evaluation and application. 
By synthesizing these existing approaches and formalizing the distinction between characteristic and descriptive labels, this work provides a framework for systematically contrasting these methods and evaluating human-like automated label generation.

\subsection*{Related approaches}
The descriptive labeling process shares similarities with two natural language processing tasks: text summarization and text classification.

Text summarization involves reducing one or more lengthy documents to a concise and readable summary that retains essential information.
Extractive summarization methods~\cite{moratanch2017extractive} are analogous to characteristic labeling in that they score passages based on their importance and concatenate the top-scoring passages to form the summary. 
However, advances in language models have enabled abstractive forms of summarization, which synthesize the source material in a way that more closely resembles what a human might produce~\cite{zhang2024summarization, liu2023summarization, pu2023summarization, liu2024learning}.
Despite these parallels, descriptive labeling differs from abstractive text summarization in several key ways.
First, whereas summarization typically produces summaries spanning multiple sentences, descriptive labeling aims to generate a concise label comprising only a few words.
Second, whereas abstractive summarization seeks to distill a text’s most pertinent information, including findings, concepts, or theories, descriptive labeling focuses on creating a coherent and readable label that broadly explains the types of documents within a cluster rather than their specific content.

At first glance, descriptive labeling also resembles text classification. 
Text classification assigns texts to one or more predefined categories within an established taxonomy. 
The key distinction is that classification relies on an \textit{a priori} set of exhaustive categories, whereas descriptive labeling operates without such an explicit scheme, allowing for an unbounded range of potential labels.
That said, taxonomies are not irrelevant to labeling. 
For a label to be human-readable, it should resemble taxonomic categories familiar to readers, such as disciplinary classifications. 
Some historical labeling methods have, in fact, made use of existing ontologies to assign labels, such as categories and pages from Wikipedia~\cite{allahyari2015ontology,Reddivari2019,carmel2009}.
Furthermore, language models have been shown to encode domain knowledge from their training data~\cite{singhal2023clinical,petroni2019knowledge}. 
In the context of scientific and library materials, such models encode taxonomic information that informs the generation of labels.

\subsection*{Formalizing the cluster labeling task}
Here, we provide a semi-formal definition of cluster labeling which will clarify the procedure and offer a shorthand for discussions of aspects of this procedure for the rest of the study.
The goal of this exercise is to make the various components of the descriptive labeling task explicit and in so doing to identify the decisions to be made when implementing a labeling workflow.

Before labeling, we assume that a corpus of scientific papers $P = [p_1, p_2, p_3, \ldots]$ has been passed to a clustering function $\text{Cluster}(P) \rightarrow C$. 
This function maps each paper, $p_i$, to membership in one of an arbitrary number of clusters $C = [c_1, c_2, c_3, \ldots]$.
For example, a cluster might resemble $c_1 = [p_{4}, p_{43}, p_{56}, p_{98}, p_{109}, \ldots]$. 
The definition of $\text{Cluster}(P)$ is a design choice, although common approaches consider citation relations~\cite{subelj2016clustering}
Here we assume that clustering is based on topical similarity between papers. 

The labeling task can be conceived as a function that maps the set of all clusters to a set of labels, $\text{GenerateLabel}(C) \rightarrow L$, where the label for a particular cluster $c_i$ is represented as $l_i$.
An alternative formulation for this function is $\text{GenerateLabel}(c_i) \rightarrow l_i$, which is a function that maps a single cluster to a single label using only local information.
This alternative formulation is simpler and likely useful for certain tasks. 
However, we argue that, for most tasks, effective labeling requires incorporating information across all clusters to inform each individual label.

At this point, the characteristic and descriptive labeling procedures diverge.
Characteristic labeling is a multi-step process that typically includes (i) extracting relevant characteristics from the content and metadata of documents in each cluster; (ii) applying some function to rank characteristics by relevance for labeling, usually based on within-cluster prominence and between-cluster distinctiveness; and (iii) selecting some number of top-ranked characteristics and forming them into a final label.
Steps (i) and (ii) could be represented as $\text{Characteristics}(C) \rightarrow F$, where $F_i$ represents the list of characteristics of documents in a cluster $i$ ordered by some ranking algorithm, such that $F_i = [f_{i,1}, f_{i,2}, f_{i,3},\ldots]$.
The implementation of $\text{Characteristics}(C)$ is a design choice, but characteristic labels will typically involve identifying important and distinctive features of the documents in each cluster.
The final step to generate a characteristic label (iii) is $\text{GenerateLabel}(F_i) \rightarrow l_i$, a function that maps the features of a particular cluster, $F_i$ to a final label, $l_i$. 
For characteristic labels, $\text{GenerateLabel}(F_i)$ will tend to take the form of concatenating some number of characteristics into a string, such that $l_i = \text{``}f_{i,1}\text{;}f_{i,2}\text{;}f_{i,3}\text{;}\ldots;f_{i,n}\text{''}$, where $n$ is the number of characteristics to select and ``;'' is a separator token to aid reading. 

Descriptive labeling with LLMs is a less well-defined task, but for this paper we consider a process similar to characteristic labeling:
the most prominent and distinctive characteristics of documents in each cluster are extracted by some function $\text{Characteristics}(C) \rightarrow F$.
However, the function for generating the final label is replaced with a prompt to a language model such as ChatGPT. 
The function for generating a label can then be represented as $\text{GenerateLabel}(F_i, \text{model}, \text{template}, \gamma) \rightarrow l_i$, for which the function parameters include the choice of language model (e.g., ChatGPT, Claude, LLaMA), the template to encode the characteristics $F_i$ into a prompt that can be submitted to a language model, and a set of language-model-specific parameters, $\gamma$ (e.g., temperature for ChatGPT).
The resulting label $l_i$, takes the form of a string generated by the language model in response to the created prompt and parameters. 
The selection of $\text{model}$, $\text{template}$, and $\gamma$ are workflow design decisions.

We note that whereas characteristic labeling follows a systematic and deterministic procedure, language models tend to introduce stochasticity into their responses~\cite{ouyang2024nondeterminism} so responses are non-deterministic such that a quality label cannot be ensured by inputs.

The use of language models to generate descriptive labels introduces two major challenges.
First, language models may produce improper responses.
These may take the form of ``hallucinations'' such that the model generates responses that are nonsensical or irrelevant to the prompt~\cite{huang2024hallucination}.
In other cases, the response may be appropriate but malformed.
For instance, our own experiments frequently observe language models responding in the form
``The label that best fits this cluster is Magnetospheric physics''.
Such responses are frequently returned even when the language model is prompted to respond with the label alone, though this has become less of an issue given recent innovations that enforce structured outputs.
This introduces the need for additional processing to extract the label from irrelevant text.
We term this as within-label validity, though advances in language models, in particular, the use of structured outputs, have ameliorated this issue. 

The second and more pertinent validity is when a language model may generate labels that are appropriate to a particular cluster, but that are not appropriate in relation to all clusters. 
To elaborate, most language-model-based models will generate labels for one cluster at a time, and there is no built-in mechanism to enforce global criteria, such as excluding duplicate labels or labels that are relevant but so vague as not uniquely identify the cluster.
We term this problem across-label validity.

These challenges highlight the necessity of a validation step in the descriptive labeling procedure to address lapses in validity. 
Specifically, we consider a function $\text{Validate}(L) \rightarrow L'$ that is given a set of labels, $L$, assesses them according to task-relevant criteria, and returns a list of alternative labels, $L'$ where $L' \subseteq L$. 
What to do with $L'$ will depend on the specifics of a labeling task.
Generally, they will be re-generated with appropriate global contextual information appended to the prompt.

Hierarchical clustering is also frequently used in bibliometric workflows.
In this circumstance, labels must make sense with regard to the hierarchical structure of the clusters.
The current paper does not specifically examine the hierarchical structure, although the method we introduce can be modified to introduce information about a parent cluster to the generation of labels for its children by including it among the cluster characteristics, $F$.

Together, this formalization identifies the major design decisions for a labeling workflow: the extraction of characteristics, $\text{Characteristics}(C)$, the choice of a label generation function, $\text{GenerateLabel}(F_i)$ that differentiates characteristic and descriptive labeling, and for descriptive labeling the choice of model parameters such as $\text{model}$, $\text{template}$, and $\gamma$, as well as the selection of the final validation function, $\text{Validate(L)}$. 

\section*{Methods}
In this section, we describe our workflow for clustering scientific documents and generating labels. 
We outline the bibliometric data upon which we base this implementation, the methods for document clusters, the characteristics we derive from the scientific documents, the approach to labeling, the design of the experiment used in this paper to assess descriptive labels and compare implementation details, and finally an evaluative framework for making these comparisons. 

\subsection*{Data}
The scientific documents used in this study were sourced from the Dimensions bibliographic database~\cite{Hook2018,Hook2021,Herzog2020}, which at the time of writing indexed more than 140 million scientific publications.
We focused our experiments on four fields that represent a broad range of disciplines and, importantly, exhibit distinct styles of technical vocabulary that may have consequences for the efficacy of descriptive labeling. 
These fields were identified using the second (“group”) level of the 2020 Australia \& New Zealand Standard Research Classification Fields of Research (FoR) codes.
Dimensions assigns FoR codes to individual publications through a classifier applied to their titles and abstracts~\cite{Hook2018,porter2023FoRcodes}. 
Each publication may be assigned up to two FoR codes.
The four fields are identified as ``Plant biology'' (FoR: 3108), ``Oncology and carcinogenesis'' (FoR: 3211), ``Artificial Intelligence'' (FoR: 4602), and ``Applied and developmental psychology'' (FoR: 5201). 
For each of these FoR codes, we searched Dimensions for relevant English-language journal articles and conference proceedings published between 2003 and 2023. 
From these, we randomly sampled 15,000 records for each field, which, from our cursory observations, adequately capture a range of granular sub-topics that can be identified via unsupervised clustering. 
These documents collectively represent our corpus of publications $P$.
For each record in $P$, we collected data on its title, venue (journal or conference name), citation impact, and ``concepts''—noun phrases extracted by Dimensions from titles and abstracts, intended to represent the fine-grained topics of the publication~\footnote{\url{https://docs.dimensions.ai/dsl/language.html}}.

\subsection*{Generating document clusters}
Prior to labeling clusters, the clusters must first be identified by defining an operation $\text{Cluster}(P) \rightarrow C$.
Although descriptive labeling is not tied to any particular clustering approach, it is still important to understand the characteristics of clusters and that clusters be of adequate size and distribution to ensure characteristics can be reliably extracted and quality labels produced. 
We implemented a two-step clustering approach. 

In the first step, we generated embedding vectors for each document in our corpus.
Vectors were generated using SPECTER~\cite{cohan-etal-2020-specter}, a fine-tuned variant of SciBERT~\cite{beltagy-etal-2019-scibert} that takes as input the title and abstract of each document and generates a dense vector representation.
In the resulting space of vectors, the distance between publication vectors roughly corresponds to their topical relatedness.

The second step involved reducing the dimensionality of the vectors.
For each of the four fields, we use UMAP~\cite{McInnes2018} to embed publication vectors in 2 dimensions.
This dimensionality reduction makes subsequent clustering more computationally efficient.
Next, we applied the unsupervised HDBScan algorithm to automatically identify clusters in the 2-dimensional space of publications. 
Together, there are several hyper-parameters that must be set for UMAP and HDBscan, the choice of which can change the quality of the resulting clusters.
Although we do not aim for perfect clusters, we seek to reduce the 15,000 publications in each field to something near 100 clusters, and while inequality in cluster size is inevitable, we aim for no clusters smaller than 20 documents.

\subsection*{Cluster characteristics}
We provide an implementation of the operation $\text{Characteristics}(C) \rightarrow F$ to identify the prominent and distinctive characteristics for each cluster. 
Although there are many possible characteristics that can be extracted from bibliometric metadata, we consider three.

\textbf{Characteristic Terms}: 
The most frequent and distinctive terms from the titles and abstracts of the cluster documents have historically been identified using a variety of approaches, such as extracting noun phrases or \textit{n}-grams and weighting them via TF-IDF~\cite{li2015labeling}, mutual information~\cite{koopman2017mutual}, temporal emergence~\cite{chen2005citespace}, or other term ranking metrics. 
In this study, we made use of the "concepts" field from the Dimensions database, which contains noun phrases extracted from document abstracts and weighted by their relevance to the respective document.
Concepts with relevance scores below a specified threshold were excluded. 
Subsequently, we computed the TF-IDF for the remaining concepts within each cluster. 
Term frequency (TF) represents the total occurrences of each concept within a cluster, whereas inverse document frequency (IDF) accounts for the distribution of concepts across all clusters. 
This calculation assigns higher importance to concepts distinctive to a cluster and penalizes those common across all documents. 
Finally, the 12 highest-ranking concepts for each cluster were selected as characteristic terms.

Although our selection of 12 concepts was initially based on trial and error in early development, we conducted additional analysis to examine how the number of concepts affects the quality of the label. 
Using the ``Label-shift'' metric described below (Table~\ref{tab:concepts_prompt_comparison}), we found that 12 concepts produced similar labels as 16 concepts, but labels became increasingly dissimilar as fewer concepts were used. 
Qualitative examination of example labels confirms that fewer concepts tended to produce less specific labels, whereas 12 concepts provide sufficient information for effective labeling for our task (Table~\ref{tab:examples_by_num_concepts}). 
The choice of number of characteristic terms should be based on task and implementation details.

\textbf{Prominent venue titles}: 
The titles of venues, such as journals and conference proceedings, are carefully selected to summarize the types of papers they publish, making them a valuable resource for labeling~\cite{velden2017journal-signature}. 
Previous studies have used the titles of the most frequent journals within a cluster as cluster labels~\cite{lamers2021disagreement}. 
In this work, we selected the names of the three most frequent venues within each cluster based on their occurrence.

\textbf{Prominent document titles}:
The titles of prominent papers within a cluster can provide valuable insight into the key topics and themes represented by the cluster.
However, there are various ways to measure the prominence of a document. 
In this study, we adopt a straightforward approach, ranking documents according to their field-citation ratio, a field-normalized citation metric provided by Dimensions. 
The three highest-ranking documents in each cluster, according to this metric, were selected as the prominent document titles. 
This approach frequently highlights review articles, seminal works, or definitive papers that effectively summarize the cluster's main themes. 
Future work could explore methods specifically aimed at identifying review articles, so-called "authoritative" papers~\cite{boyack2020comparison}, or documents most representative of the cluster as a whole.

The primary mode of interaction with language models is through natural language. 
Therefore, once cluster characteristics are identified, they must be encoded into a textual prompt suitable for submission to the language model. 
Prompts were composed of two main components: the \textit{template}, which provides a preamble with task-specific instructions for the language model, and the \textit{clauses}, which convert cluster characteristics into natural language phrases included in the final prompt. 
Each characteristic was formatted as a comma-separated list. 
Table~\ref{table:prompt-description} illustrates an example of a prompt description and its associated clauses as implemented in our approach.

\begin{table}[p]
\renewcommand{\arraystretch}{2.0}
\setlength{\tabcolsep}{12pt}
\caption{
\textbf{Prompt template and clauses.}
Example of template components for a minimal prompt. 
The ``Template'' row specifies the prompt template that describes the task to which the language model should respond. 
Clauses are inserted into the template in the marked location. 
Clauses encode the cluster characteristics into the prompt in the form of a comma-separated list inserted into the location marked ``\{...\}'' following the clause description.
In addition to the three characteristic clauses, shown are also the ``Duplicate'' and ``Non-specific'' clauses that are used in the validation step.
}
\label{table:prompt-description}
\small
\resizebox{\textwidth}{!}{%
\begin{tabular}{p{1.25in}|p{4in}}
\textbf{Template} &
    {\par
        Generate a label for the scientific specialty represented with the following information extracted from a cluster of related documents. 
        {\color{orange} \{clauses go here…\}}. 
        The label should resemble something that is already present  in existing ontologies. The label should be as specific as possible while still representing all of the provided information. Additionally, the label should be short and not use any redundant words.
    }
    \\
\textbf{Characteristic terms clause}  & 
    The concepts most associated with these documents in order from most to least relevant are: {\color{blue}\{...\}} \\ 
\textbf{Prominent journals clause} & 
    Most documents come from journals such as {\color{ForestGreen} \{...\}} \\
\textbf{Prominent papers clause} & 
    The most prominent articles in this cluster are titled {\color{Mulberry}\{...\}} \\
\textbf{Format clause} &
    The following labels are invalid and should not be used: \{...\}. \\
\textbf{Duplicate clause} & 
    The label should not be any of the following: \{...\}. \\
\textbf{Non-specific clause} & 
    The new label should be different from and more specific than: \{...\}.
\end{tabular}%
}
\end{table}

\subsection*{Label validation}
\begin{figure}[h!]
    \centering
    \includegraphics[width=0.80\linewidth]{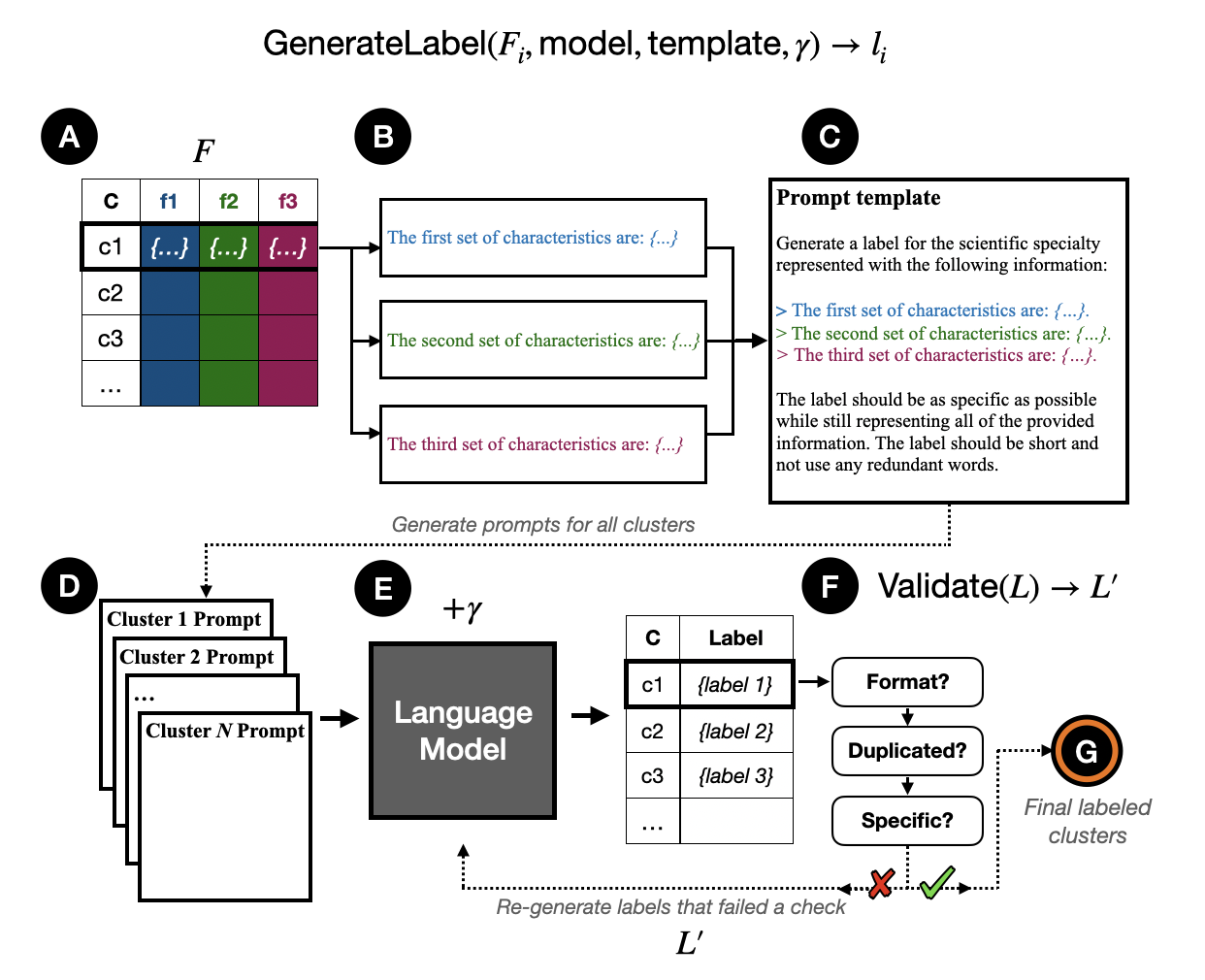}
    \caption{
    \textbf{Diagram of our general approach to descriptive label generation.} 
    Illustrates the operation $\text{GenerateLabel}(F_i, \text{model}, \text{template}, \gamma) \rightarrow l_i$.
    Assume that there exists a set of papers, $P$ that have been mapped to clusters $C = [c_1, c_2, c_3, \ldots]$, and that from these clusters prominent characteristics have already been surmised.
    \textbf{(A)} begin with $F$, which lists the top characteristics for each cluster.
    \textbf{(B)-C} Defines a prompt template, and the process by which characteristics are encoded in the template through what we term ``clauses''. 
    \textbf{(D)} A prompt is created for each cluster. 
    \textbf{(E)} For each prompt, a language model is queried and output labels collected.
    The query includes additional model-specific parameters, $\gamma$.
    \textbf{(F)} The core of the iterative labeling approach. 
    Each label is assessed based on certain checks, representing the operation $\text{Validate}(L) \rightarrow L'$.
    Here, specific validation criteria include whether the label is locally valid (e.g., were instructions followed), whether it is duplicated (appearing multiple times), and whether the label is appropriately specific. 
    Labels that fail this validation are represented as $L'$, and are re-generated until all labels pass validation.
    \textbf{(G)} The result of this procedure is the final labeled set of clusters, $L$.
    }
    \label{fig:general-approach}
\end{figure}

Our general approach to implementing the operation $\text{GenerateLabel}(F_i, \text{model}, \text{template}, \gamma) \rightarrow l_i$ is shown in the diagram in Fig.~\ref{fig:general-approach}.
We adopt an iterative approach to label generation,  enabling the validation of language model responses and the enforcement of across-label validity.
Our approach first assigns an initial label to each cluster. 
Then, a loop is initiated, during each iteration we execute the operation $\text{Validate}(L) \rightarrow L'$.
Although many such validations could be implemented, for the sake of the present paper, we include three:

\begin{itemize}
    \item \textbf{Format}: we determine whether the label is valid based on its length, between 3 and 50 characters; too long labels signal a failure or an overly verbose response from the language model.
    \item \textbf{Duplicated}: we examine whether a label is a duplicate, meaning that the same label appears more than once among the labeled clusters. 
    \item \textbf{Non-specific}: we evaluate the specificity of a label. This criterion ensures that a label is not overly vague, such that it could plausibly describe many different clusters. Specificity is assessed by comparing the vector representation of the label (generated using OpenAI’s embedding client) to the vector representation of a sentence summarizing the cluster’s distinguishing characteristics.
    The comparison is performed using cosine similarity, where higher values indicate greater semantic alignment. 
    A label is considered sufficiently specific if its similarity to its corresponding cluster is higher than its similarity to any other cluster.
\end{itemize}

Labels that fail any of these checks are immediately re-generated. 
To address issues with the previous label, the prompt is updated with an additional clause describing the specific problem and encoding the problematic label as an example to avoid in subsequent generations (see Table~\ref{table:prompt-description} for details of the clause descriptions). 
For the "Duplicated" clause, instead of retaining one duplicate, all duplicated labels are re-generated. 
This validation procedure is repeated until no labels require re-generation ($|L'| = 0$) or a maximum of 10 iterations is reached in our implementation.
If the maximum is reached, the most recent labels are retained.
Although these checks are not exhaustive of all possible or useful validations, we found that, in practice, they reliably produced high-quality labels.

\subsection*{Experimental Design}
Using the data and implementation of the descriptive labeling procedure defined previously in this section, we design a series of experiments to address three key questions.
First, to what extent do the choices in the design of a clustering actually matter? 
The choice of language model and prompt design may have consequences for the generated labels.
We aim to assess which choices are important in order to guide future work in label generation. 
Second, are descriptive labels comparable to characteristic labels in their ability to distinguish clusters?
We consider descriptive labels to be \textit{prima facie} preferable to characteristic labels due to their aesthetics.
However, we also aim to examine their performance relative to characteristic labels and, in doing so, provide evidence to support their use in future analyses and applications.
Third, to what extent do results differ by field?
Disciplines differ in their vocabulary and writing conventions, which may have consequences for their ability to generate effective disciplinary labels.
By examining our four disciplinary cases, we aim to understand whether such techniques work better for some disciplines than others.

We do not consider the experiments outlined here to be exhaustive.
The design choices in a labeling system are practically limitless.
Instead, we aim to provide a preliminary response to our three major questions while also establishing a framework for descriptive label evaluation that may be extended in future work.

Of the publicly available language models, we focus on those produced by OpenAI and collectively referred to as "ChatGPT". 
Claude, LLaMA, and Geimini are also available, each of which is distinct in terms of training and fine-tuning.
Although we initially planned to compare all models, we quickly found that they tended to produce malformed responses, whereas the ChatGPT models reliably produced valid results. 
Modifications to the prompt, such as requesting responses in the form of a JSON object, may ameliorate these issues; however, they were not implemented for the current study. 
Our focus on the OpenAI ecosystem allows us to contrast models that differ in cost and size to determine the minimum level of sophistication necessary to perform labeling tasks, which can inform the choice of model more generally. 
We examine three of the ChatGPT models: ChatGPT3 (\textit{gpt-3.5-turbo-0125}), ChatgGPT4-mini (\textit{gpt-4o-mini-2024-07-18}), ChatGPT4 (\textit{gpt-4o-2024-05-13}), and ChatGPT4o (\textit{gpt-4o-2024-11-20}).

This study evaluates two types of prompt. 
The first is a minimal version, as described in Table~\ref{table:prompt-description}. 
The second builds upon the first by adding a "system prompt," commonly used in OpenAI models to define a specific "role" for the language model to emulate~\cite{shanahan2023roleplay}. 
In practical terms, the system prompt serves as a preamble injected at the beginning of the prompt. 
The goal of this comparison is to determine whether the inclusion of a system prompt results in significant changes in the generated labels. 
In our implementation, the system prompt instructs the language model to act as a librarian tasked with classifying documents.

\begin{quote}
    ``You are a librarian with an expertise in the taxonomy of knowledge and scholarship. Your job is to examine characteristics about clusters of scientific publications and to assign an appropriate label.''
\end{quote}

Finally, we examine the validation step, $\text{Validate}(L) \rightarrow L'$, which can significantly increase execution time due to the interaction with language model APIs. 
This delay arises primarily from two factors: the need to re-generate labels, which requires additional API requests, and the \textit{Non-Specific} validation check.
The latter involves creating vector representations of labels and characteristics using the OpenAI embedding API, a process that is computationally intensive. 
Although we implement vector caching to minimize API calls, this step can still be time-consuming. 
To evaluate the impact of validation, we consider two scenarios. 
The first involves label generation without any validation step, allowing us to assess the necessity of validation. 
The second includes full validation with all checks described in this paper.

For every combination of experiment parameters, we execute 10 runs of the workflow, producing 10 sets of labels.
Most language models implement a degree of stochasticity into their outputs, and so by executing 10 runs, we are able to assess a baseline of variance that is useful for downstream quantitative evaluation.

\subsection*{Quantitative evaluation metrics}
We propose a metric called ``\textit{label-shift}'', which aims to provide a means of quantitative assessment of workflow decisions.
Given two sets of generated labels, $L_i$ and $L_j$ generated through different workflow designs (e.g., different prompts), label-shift provides a single value quantifying the extent of their semantic difference.
This is not a measure of the quality of a label set, but rather a means of comparing label sets.
Here, label-shift allows us to compare label sets produced through different workflows that vary by their model or prompt template.

To compute label-shift, we generate a vector representation for each value of each set, $\text{GenerateVectors}(L_i = [l_{i,1}, l_{i,2}, l_{i,3}\ldots]) = V_i = [v_{i,1}, v_{i,2}, v_{i,3} \ldots] $, where \textit{GenerateVectors} is a function that produces a vector representation of a given label. 
Then we define a measure of similarity for every pair of vectors, $\text{\textit{sim}}(v_{i,z},v_{j,z})$, where $\text{\textit{sim}}$ is between 0 and 1 (with 1 being most similar), and $z$ designates the index of the cluster for which the label (and vector) were generated. 
Label-shift, then, is defined as the average similarity across all vector pairs
$\text{\textit{LS}} = \frac{\sum_{z = 1}^{|L_i|} \text{sim}(v_{i,z}, v_{j,z})}{|L_i|}$, 
where $|L_i|=|L_j|$, where \textit{LS} is between 0 and 1, with 1 indicating the highest similarity. 

This formalization reveals several choices that can be made to calculate LS. 
First, at its most basic, \textit{GenerateVectors} could produce a one-hot encoded D-dimensional vector where $D = |L_i \cup L_j|$------the total number of unique labels across relevant label sets.
However, one-hot vectors only allow for exact matching and don't capture the semantic similarity between labels; for example, ``black hole physics'' and ``physics of black holes'' would be just as dissimilar as ``black hole physics'' as ``molecular biology''. 
Here, we draw on the variety of language model approaches for generating dense and continuous vector-space representations of arbitrary texts. 
Although options exist for scientific text (e.g., SciBert~\cite{beltagy-etal-2019-scibert}), we opt for the convenience of the embedding API provided by the OpenAI API.
Then, for $\text{\textit{sim}}$ we use cosine similarity.

By itself, label-shift can be difficult to interpret.
In order to provide a basis for interpretation, we take advantage of the 10 runs of each workflow used in our experiment to calculate a two-sample Z-score that compares the distribution of label-shift values between two label sets.

We also consider a second family of metrics, which we term ``first-pass metrics''.
These metrics leverage the validation checks in our iterative labeling approach by counting the number of generated labels (before applying checks or iterations) that fail the validation criteria when the validation step is not applied.
The rationale is that different prompts and input information can elicit higher-quality responses initially, reducing the need for extensive validation and iteration.
When paired with label-shift, these metrics can also provide some context on why certain design choices impact results.

\subsection*{Manual evaluation}
Our goal is to evaluate whether descriptive labels, while \textit{prima facie} more appealing than characteristic labels, offer equal or better utility in identifying clusters. 
A key challenge in designing this evaluation lies in the fundamental differences between characteristic and descriptive labels, which make direct side-by-side comparisons uninformative. 
For instance, asking human raters to choose the better label between these two types is unlikely to yield meaningful insights. 
Instead, we design an evaluation task that separates characteristic and descriptive labels, enabling a more objective comparison. 
Specifically, we focus on the ability of a label to unambiguously map to its corresponding cluster.
Characteristic labels were assigned to each cluster consisting of the top 12 characteristic ``concepts'' that were used in the descriptive labeling task, with each term separated by a semicolon.

We model the annotation task as a multiple choice quiz, in which an annotator is presented with a list of prominent journals and papers that appear in the cluster (the same that are used to determine the label) and a list of four possible cluster labels, one of which is the correct label that was assigned to the cluster and the other three are randomly selected from among clusters in the same dataset. 
Characteristic terms are not shown as the presented labels will be either these terms shown as a characteristic label, or the descriptive label. 
The annotator is tasked with selecting the label that best represents the cluster based on the information shown.

The annotation task was repeated for a random sample of 50 clusters across each of the four disciplinary datasets.
Given the effort involved, we could not annotate all combinations of possible implementation details (e.g., across models, parameters, prompts). Therefore, we focus on descriptive labels produced using ChatGPT4, the prompt that includes a ``system prompt'', and with all validity checks enabled. 
Annotators repeated the task across two scenarios: one in which only descriptive labels were shown and the other in which only characteristic labels were shown. 

Annotators were drawn from the authors of this paper who have a sufficiently high level of general scientific knowledge to inform their decisions. 
However, we note the difficulty of this evaluation given the jargon and technical information involved.
To assess accuracy, we evaluated inter-annotator reliability.
Two annotators each selected labels for the entire ``Botany'' dataset of descriptive labels of 50 clusters, of which there was agreement on 38 (76\%) of labels, which we believe to be a sufficiently high level of agreement given the difficulty of the task. 

%
%
\section*{Results}
\subsection*{Label-shift}
The label-shift metric is used to understand the extent to which prompt characteristics lead to semantically distinct labels. 
We compare the distribution of semantic similarity scores between runs of the same baseline prompt against the mean similarity between the baseline and an alternative prompt (Table~\ref{table:shift_prompt_comparison}). 
The result is a Z-score that quantifies the semantic difference between the labels produced by the baseline and alternative prompts given the typical variation between runs. 
Because similarities are likely lower than the baseline variation, this comparison can be conceived as a one-sided left-tailed test, and so we opt to use the conventional Z-score critical value of -1.645 as a rough guide for interpretation. 
The baseline prompt includes all aspects of the template (Table~\ref{table:prompt-description}), including a system prompt, the concept clause, the paper clause, and the journal clause; alternative prompts drop one or more of these.
All comparisons are made against the baseline prompt. 
We note that a difference with the baseline does not signal that the alternative labels are better or worse, merely that the labels produced are distinct.

Across each of the four fields of study, we observe that the absence of the system prompt had the lowest effect on the semantic similarity of labels (Table~\ref{table:shift_prompt_comparison}, ``Concepts, Papers, Journals'').
The exclusion of journals also had minimal consequence.
Using solely the concepts clause (excluding papers and journals but keeping the system prompt) did lead to semantically distinct labels, though the difference did not rise to the critical value of -1.645 that we use to guide interpretation.
The use of only the top papers clause (excluding concepts, journals, but keeping the system prompt) had the largest difference across all four disciplines, with the strongest reported for the field of Artificial intelligence ($Z = -2.172$) and even at its lowest ($Z = -1.600$, ``Oncology and carcinogenesis'') it approached the critical threshold. 
This result signals that using only the titles of the top papers leads to considerably different results from the baseline.
Examining the first-pass evaluation results (Table~\ref{table:prompt_eval_firstpass}), we see that when using only the papers clause, the number of results identified as too vague was greater than the baseline for each dataset.
This suggests that the titles of top papers do not provide sufficient information to create sufficiently precise labels, and although our labeling workflow re-generates vague labels, the results end up semantically distinct from the baseline.

Next, we examine the cluster labels generated using only titles that were flagged as too vague, comparing them to labels created using all characteristics (Table~\ref{si:tab:title_examples}). In most cases, poor labels resulted from titles that were not representative of the broader cluster content. For instance, the highest-cited papers in Oncology \& carcinogenesis cluster \#8 (\#3211) lack cardiology-related terms that are prevalent throughout the cluster when revealed by distinguishing concepts. This problem may be amplified when field-level citation differences elevate topically peripheral papers to the top of clusters.

\begin{longtable}{lrrrr}
\caption{
\textbf{Label-shift comparison between baseline and alternative prompts.}
The baseline includes all features, including a system prompt, top concepts, top papers, and most frequent journals. 
Other prompts exclude one or more of these characteristics.
Labeling was performed using ChatGPT4 and repeated across 10 runs.
``Mean'' and ``std`` shows the average and standard deviation of the cosine similarity between the vector representations of the labels generated using the baseline and alternative prompt.
The ``Z score'' was calculated from these values; a low Z score signals that the alternative prompt produces, on average, semantically distinct labels from the baseline prompt.
} 
\label{table:shift_prompt_comparison} \\
\toprule
Prompt & Mean & std & Z score \\
\midrule
\endfirsthead
\caption[]{Comparison of different models} \\
\toprule
Prompt & Mean & std & Z score \\
\midrule
\endhead
\midrule
\multicolumn{4}{r}{Continued on next page} \\
\midrule
\endfoot
\bottomrule
\endlastfoot
\textbf{Plant Biology} & & & \\
Concepts, Papers, Journals & 0.969 & 0.002 & -0.270 \\
Concepts, Papers & 0.966 & 0.002 & -0.375 \\
Concepts & 0.952 & 0.002 & -0.860 \\
Papers & 0.920 & 0.001 & -2.017 \\
\textbf{Oncology and carcinogenesis} & & & \\
Concepts, Papers, Journals & 0.969 & 0.002 & -0.277 \\
Concepts, Papers & 0.967 & 0.002 & -0.322 \\
Concepts & 0.955 & 0.002 & -0.727 \\
Papers & 0.924 & 0.002 & -1.600 \\
\textbf{Artificial intelligence} & & & \\
Concepts, Papers, Journals & 0.973 & 0.002 & -0.283 \\
Concepts, Papers & 0.970 & 0.002 & -0.400 \\
Concepts & 0.958 & 0.002 & -0.848 \\
Papers & 0.923 & 0.002 & -2.172 \\
\textbf{Applied and developmental psychology} & & & \\
Concepts, Papers, Journals & 0.969 & 0.002 & -0.277 \\
Concepts, Papers & 0.965 & 0.001 & -0.475 \\
Concepts & 0.948 & 0.002 & -1.032 \\
Papers & 0.917 & 0.002 & -2.089 \\
\end{longtable}

We further examine whether the choice of model relates to differences in the generated labels. 
Our experiment focused on a subset of models within the OpenAI ecosystem, ranging from newer (and more expensive) to less recent but cheaper to use models. 
The results (Table~\ref{table:shift_model_comparison}) show a gradient of increasing difference while moving towards older models;
that is, the largest differences are observed when comparing ChatGPT3 against the baseline of ChatGPT4o, with more moderate differences for ChatGPT4 and ChatGPT4-mini.
Examining Table~\ref{table:model_eval_firstpass}, we observe that in all but one dataset, labels produced by ChatGPT3 were more often identified as vague, on average, though the total rate was low, fewer than one vague label per run. This suggests that the generation of overly vague labels was likely not a driver of this difference.
We also note that the Z-score between ChatGPT4o and ChatGPT3 does not reach the critical threshold of 1.645 for any dataset, hitting at closest -1.288 for ``Oncology and carcinogenesis''; although there are clear differences between label sets, we would not consider them major between the models.
Together, these results suggest that the choice of model does matter, with clear---though perhaps not necessarily substantial---differences between newer and older models.

\begin{longtable}{lrrrr}
\caption{
\textbf{Label-shift comparison between baseline and alternative language models.}
The baseline model is ChatGPT4o, which was selected because it was one of the most advanced models at the time this experiment was executed. 
All labels were generated based on the full prompt with all clauses and a system prompt. 
``Mean'' and ``std``show the average and standard deviation of the cosine similarity between the vector representations of the labels generated using the baseline and alternative prompt.
The ``Z score'' is calculated from these values; 
a low Z-score signals that the alternative prompt produces, on average, semantically distinct labels from the baseline prompt.
} 
\label{table:shift_model_comparison} \\
\toprule
Model & Mean & std & Z score \\
\midrule
\endfirsthead
\caption[]{Comparison of different models} \\
\toprule
Model & Mean & std & Z score \\
\midrule
\endhead
\midrule
\multicolumn{5}{r}{Continued on next page} \\
\midrule
\endfoot
\bottomrule
\endlastfoot
\textbf{Plant Biology} & & & \\
ChatGPT4 & 0.969 & 0.002 & -0.382 \\
ChatGPT4-mini & 0.956 & 0.002 & -0.839 \\
ChatGPT3 & 0.947 & 0.002 & -1.109 \\
\textbf{Oncology and carcinogenesis} & & & \\
ChatGPT4 & 0.968 & 0.003 & -0.373 \\
ChatGPT4-mini & 0.957 & 0.003 & -0.732 \\
ChatGPT3 & 0.949 & 0.001 & -1.288 \\
\textbf{Artificial intelligence} & & & \\
ChatGPT4 & 0.975 & 0.002 & -0.272\\
ChatGPT4-mini & 0.966 & 0.002 & -0.564 \\
ChatGPT3 & 0.953 & 0.001 & -1.103 \\
\textbf{Applied and developmental psychology} & & & \\
ChatGPT4 & 0.966 & 0.002 & -0.401 \\
ChatGPT4-mini & 0.953 & 0.002 & -0.799 \\
ChatGPT3 & 0.947 & 0.002 & -1.094 \\
\end{longtable}

\subsection*{Manual Evaluation}
To assess the quality of the labels produced by descriptive versus characteristic labeling approaches, we examine the ability of a user to uniquely identify the correct label for a set of representative papers from a list of random alternative labels.
50 clusters were annotated according to this task, twice for each characteristic and descriptive label. 
Descriptive labels are taken to be \textit{prima facie} preferable to characteristic labels in terms of legibility, but similar performance in this task would provide further evidence of their utility.

The results show that descriptive labels perform at or close to that of characteristic labels (Table~\ref{table:evaluation_results}).
Descriptive labels \textit{outperformed} characteristic labels for the field of ``Plant biology'' (78\% vs. 70\%).
Descriptive labels performed roughly equal to characteristic labels for the field of ``Artificial intelligence'' (90\% vs. 92\%).
Finally, descriptive labels had lower but still similar performance for ``Oncology and carcinogenesis'' (82\% vs. 90\%) and ``Applied and developmental psychology'' (82\% vs. 94\%).
We interpret these results as evidence that large language models are able to produce labels that uniquely identify a cluster, at least as compared to characteristic labels.

\begin{table}[ht]
\centering
\caption{
\textbf{Descriptive labels perform at or near characteristic labels at distinguishing clusters.}
For each of the four datasets, we show the percentage of the 50 clusters for which the annotator-selected label corresponds to the correct cluster.
We find that descriptive labels have stronger performance for Plant biology, near equal performance for Artificial intelligence, and slightly lower performance for ``Oncology and carcinogenesis'' and ``Applied developmental psychology''. 
}
\label{table:evaluation_results}
\begin{tabular}{lrr}
  \hline
Dataset & Characteristic & Descriptive \\ 
  \hline
Plant biology & 0.70 & 0.78 \\ 
  Oncology and carcinogenesis & 0.90 & 0.82 \\ 
  Artificial intelligence & 0.92 & 0.90 \\ 
  Applied and developmental psychology & 0.94 & 0.82 \\ 
   \hline
\end{tabular}
\end{table}

We also provide further qualitative insights about the annotators' experiences with this task that contextualize these results. 
First, although we did not measure the time taken to select each label, the annotators reported that selecting the correct characteristic label was time-consuming compared to selecting descriptive labels.
Second, the annotators also reported distinct strategies to identify the correct label for each type.
That is, for descriptive labels, the label was understood to be summative and represented a topic, which was then used to place the representative papers.
For characteristic labels, however, jargon dominated the label, leading annotators to rely primarily on keyword matching --- identifying common terms shared between the label and the papers --- rather than forming a holistic understanding of the topic. 

For example, consider the following set of papers and corresponding characteristic and descriptive labels from plant biology (cluster 54),

\begin{itemize}[topsep=0pt,itemsep=-1ex,partopsep=1ex,parsep=1ex]
    \item \textbf{Paper 1:} ``The Mg-chelatase H subunit is an abscisic acid receptor''
    \item \textbf{Paper 2:} ``Low phosphate activates STOP1-ALMT1 to rapidly inhibit root cell elongation''
    \item \textbf{Paper 3:} ``Abscisic acid signaling and crosstalk with phytohormones in regulation of environmental stress responses''
    \item \textbf{Characteristic Label:} ``abscisic acid; Abscisic; dormancy; seed dormancy; plant hormone abscisic acid; phytochrome-mediated responses; Arabidopsis; tertiary structure; abscisic acid homeostasis; hypocotyl elongation; mutants; hormone abscisic acid''
    \item \textbf{Descriptive label:} ``Plant Abscisic Acid Regulation and Dormancy Mechanisms''
\end{itemize}

For these papers, the characteristic label was identified by matching the key term ``abscisic'' to its use in the titles of two representative papers.
The same approach facilitated the selection of the descriptive label, but the label is more legible and provides a sense of the topic even for those with little knowledge of plant biology.
This suggests that the performance of characteristic labels is partially the result of this keyword-matching approach, rather than the label's ability to adequately describe the topic of the cluster.

%
\section*{Conclusion}
The results of our manual evaluation show that descriptive labels, in addition to being \textit{prima facie} more legible and human-interpretable than characteristic labels, perform at or near the same level in uniquely identifying clusters. 
This provides an empirical basis for their continued use in bibliometric workflows and support for the proliferation of LLM-based labeling across the field of bibliometrics.

Our analysis shows that the most influential factors in an LLM-based labeling prompt are the inclusion of key terms of the cluster, followed by the titles of representative papers. 
Titles of highly-cited papers may fail to provide reliable cluster labels because these papers are often unrepresentative of their clusters. 
Alternative approaches, such as identifying the most topically-representative papers, may prove more effective, but are also more methodologically challenging. 
However, distinguishing concepts---key words and phrases extracted from titles and abstracts---consistently generated robust and distinct labels.
Model choice also plays a role, with older and simpler models generating labels that are somewhat semantically distinct from those produced by newer and larger models. 
Users must weigh the trade-offs between model cost and label quality; in many applications, the lower-cost ChatGPT-3 may be sufficient for label generation.

Despite its widespread adoption, LLM-based descriptive labeling largely relies on craft knowledge developed and shared informally by individual researchers. 
Often, this knowledge is proprietary and undocumented.
The rapid adoption of these approaches across major industry platforms, including large-scale implementations that far surpass our empirical evaluation in scope, underscores both their practical value and the need for transparent methodological foundations.
By providing an open, systematic framework for LLM-based cluster labeling, this work aims to support more principled adoption of these techniques and enable researchers to build upon established foundations rather than developing ad-hoc solutions.
Our empirical results ground the use of LLMs for cluster labeling, and our formalization of the descriptive labeling task clarifies design decisions and provides a foundation for further methodological development.

One limitation of this study is its scale. 
The space of possible language models, parameters, prompts, features, and datasets is vast, yet we examine only a small subset and manually validate just one.
Although commercial deployments operate on much larger scales, our systematic framework provides the theoretical grounding and transparency often absent from proprietary implementations.
A second limitation is that the quality of input clusters, including their topical cohesion, separation, and size, may influence the performance of our label generation method, as the quality of the clustering arguably sets an upper bound on the quality of the subsequent label. 
In addition, we focus only on clusters that represent narrow topical communities identified among papers from a single discipline, which is not representative of all clustering workflows.
Future research should investigate how different clustering approaches and parameters affect the downstream labeling process.
Despite these constraints, our study offers key insights that can guide the use of language models in bibliometrics. 
The framework we present enables future researchers to systematically explore alternative configurations and evaluate their own descriptive labeling approaches.

%
%
%
%
\subsection*{Acknowledgments}
D.M. is grateful for support from the NSF (Grant \#2219575).
We thank the discussions of colleagues at Digital Science.
For their helpful comments and contributions, we thank Spoorthi Veeresh and Aquila Peeran.

\subsection*{Author Contributions}
D.M, T.H. designed research;
D.M., T.H., and C.N., W.G. performed research; 
D.M. analyzed data; 
D.M., T.H., W.G., and C.N. wrote the paper.

\subsection*{Declaration of competing interests}
T.H. is a current employee and D.M. a former employee of \textit{Digital Science}, the company that produced the bibliographic database Dimensions.
The remaining authors declare no competing interests.

\subsection*{Data availability statement}
Bibliometric data on papers was sourced from Dimensions; extracted features and annotated clusters are available alongside this manuscript.
All data and code associated with this project can be found on Zenodo at \url{https://doi.org/10.5281/zenodo.15230044}.

\clearpage

\bibliographystyle{apacite}
\bibliography{ref.bib}{}

\clearpage
\section*{Appendix}

\begin{longtable}{lrr}
\caption{
\textbf{First pass evaluation of prompt alternatives.}
Shown is the average number of duplicate and vague labels generated per run of each of five prompts.
``Baseline'' includes all aspects of the template described in Table~\ref{table:prompt-description}, including the system prompt and all clauses.
Alternative prompts exclude one or more of these aspects.
}
\label{table:prompt_eval_firstpass} \\
\toprule
Prompt & Duplicate & Vague \\
\midrule
\endfirsthead
\caption[]{...} \\
\toprule
Prompt & Duplicate & Vague \\
\midrule
\endhead
\midrule
\multicolumn{3}{r}{Continued on next page} \\
\midrule
\endfoot
\bottomrule
\endlastfoot
\textbf{Plant biology}  & & \\
Baseline & 0.100 & 0.100 \\
Concepts, Papers, Journals & 0.200 & 0.050 \\
Concepts, Papers & 0.000 & 0.050 \\
Concepts & 0.100 & 0.100 \\
Papers & 0.600 & 6.400 \\
\textbf{Oncology and carcinogenesis} & & \\
Baseline & 0.100 & 0.000 \\
Concepts, Papers, Journals & 0.100 & 0.000 \\
Concepts, Papers & 0.000 & 0.000 \\
Concepts & 0.000 & 0.000 \\
Papers & 0.000 & 3.450 \\
\textbf{Artificial intelligence} & & \\
Baseline & 0.000 & 0.000 \\
Concepts, Papers, Journals & 0.000 & 0.050 \\
Concepts, Papers & 0.000 & 0.000 \\
Concepts & 0.000 & 0.000 \\
Papers & 0.300 & 3.700 \\
\textbf{Applied and developmental psychology} & & \\
Baseline & 0.100 & 0.250 \\
Concepts, Papers, Journals & 0.000 & 0.000 \\
Concepts, Papers & 0.000 & 0.000 \\
Concepts & 0.000 & 0.000 \\
Papers & 0.000 & 2.250 \\
\end{longtable}

\begin{longtable}{lrr}
\caption{
\textbf{First pass evaluation of prompt alternatives.}
Shown is the average number of duplicate and vague labels generated per run of each of five prompts.
``Baseline'' includes all aspects of the template described in Table~\ref{table:prompt-description}, including the system prompt and all clauses.
Alternative prompts exclude one or more of these aspects.
}
\label{table:model_eval_firstpass} \\
\toprule
Model & Duplicate & Vague \\
\midrule
\endfirsthead
\caption[]{...} \\
\toprule
Model & Duplicate & Vague \\
\midrule
\endhead
\midrule
\multicolumn{3}{r}{Continued on next page} \\
\midrule
\endfoot
\bottomrule
\endlastfoot
\textbf{Plant biology} & & \\
ChatGPT4o & 0.400 & 0.150 \\
ChatGPT4 & 0.100 & 0.100 \\
ChatGPT4-mini & 0.000 & 0.350 \\
ChatGPT3 & 0.400 & 0.550 \\
\textbf{Oncology and carcinogenesis}  & & \\
ChatGPT4o & 0.000 & 0.350 \\
ChatGPT4 & 0.100 & 0.000 \\
ChatGPT4-mini & 0.600 & 0.450 \\
ChatGPT3 & 0.300 & 0.300 \\
\textbf{Artificial intelligence}  & & \\
ChatGPT4o & 0.000 & 0.050 \\
ChatGPT4 & 0.000 & 0.000 \\
ChatGPT4-mini & 0.100 & 0.000 \\
ChatGPT3 & 0.300 & 0.300 \\
\textbf{Applied and developmental psychology}  & & \\
ChatGPT4o & 0.900 & 0.050 \\
ChatGPT4 & 0.100 & 0.250 \\
ChatGPT4-mini & 1.000 & 0.000 \\
ChatGPT3 & 0.700 & 0.300 \\
\end{longtable}

\begin{longtable}{lrrr}
\caption{
\textbf{Label-shift comparison by number of concepts.}
Only top concepts, and no other cluster features, are provided in the prompt.
The baseline represents a prompt with 16 top concepts, against which other prompts with fewer concepts are compared.
Labeling was performed using ChatGPT4 and repeated across 5 runs.
``Mean'' and ``std`` shows the average and standard deviation of the cosine similarity between the vector representations of labels generated using the baseline and alternative prompt.
The ``Z score'' is calculated from these values; 
a low Z-score signals that the alternative prompt produces, on average, semantically distinct labels from the baseline prompt.
} 
\label{tab:concepts_prompt_comparison} \\
\toprule
Prompt & Mean & std & Z score \\
\midrule
\endfirsthead
\caption[]{Comparison of different models} \\
\toprule
dataset & mean & std & Z score \\
\midrule
\endhead
\midrule
\multicolumn{4}{r}{Continued on next page} \\
\midrule
\endfoot
\bottomrule
\endlastfoot
Baseline & 0.983 & 0.001 & 0.000 \\
12 Concepts & 0.963 & 0.001 & -0.981 \\
8 Concepts & 0.947 & 0.002 & -1.337 \\
4 Concepts & 0.926 & 0.002 & -1.992 \\
1 Concept & 0.874 & 0.001 & -4.617 \\
\end{longtable}

\begin{table}[]
\caption{
\textbf{Example cluster labels by number of top concepts used in prompt.}
We generate labels for all clusters in the ``Plant biology'' dataset, varied by the number of concepts included in the prompt; no other information is provided in the prompt.
Four clusters are selected as exemplars.
For these, this table shows the generated label, as well as the concepts ``added'' at each increase in the number of concepts.
For example, for cluster 45, when \#concepts = 4, that means that the concepts include ``expression'', ``salt stress'', and ``transcription factors'', in addition to the  ``Arabidopsis'' concept used when \#concepts = 1.
}
\label{tab:examples_by_num_concepts}
\resizebox{\textwidth}{!}{%
\begin{tabular}{@{}llrl@{}}
\toprule
\textbf{ID} &
  \textbf{Label} &
  \multicolumn{1}{l}{\textbf{\# concepts}} &
  \textbf{Concepts added} \\ \midrule
45 &
  Arabidopsis Genetics &
  1 &
  Arabidopsis \\
45 & Arabidopsis Salt Stress Response Mechanisms                      & 4  & expression, salt stress, transcription factors                 \\
45 &
  Arabidopsis Stress Response Transcriptional Regulation &
  8 &
  stress,transcription,mutants,genes \\
45 &
  Arabidopsis Salt Stress Response Mechanisms &
  12 &
  stress response, salt tolerance,thaliana,salt \\
45 & Arabidopsis Salt Stress Response Mechanisms                      & 16 & overexpression,Arabidopsis thaliana,expression   patterns,biosynthesis     \\
 &
   &
   &
   \\
15 &
  Pathogen Biology &
  1 &
  pathogens \\
15 &
  Pathogen Virulence Mechanisms &
  4 &
  expression, defense, virulence \\
15 &
  Host-Pathogen Interaction Genetics &
  8 &
  mutants, effector, immunity,genes \\
15 &
  Plant Immune Response to Pathogens &
  12 &
  infection, Arabidopsis, defense responses, kinase \\
15 &
  Plant-Pathogen Interaction Mechanisms &
  16 &
  resistance, immune response, cell death, host \\
 &
   &
   &
   \\
24 &
  Microbial Biocontrol &
  1 &
  biocontrol \\
24 &
  Trichoderma-Based Biocontrol &
  4 &
  Trichoderma, biocontrol agents, antifungal activity \\
24 &
  Biocontrol Agents in Fungal Pathogen Management &
  8 &
  Fusarium, endophytes, pathogens, fungicides \\
24 &
  Trichoderma Biocontrol of Fungal Pathogens &
  12 &
  Rhizoctonia, solani, oxysporum, fungi \\
24 & Trichoderma as Biocontrol Agents Against Plant Pathogenic Fungi  & 16 & Fusarium oxysporum, antagonistic activity, mycelial growth, fungus            \\
 &
   &
   &
   \\
3 &
  Data Extraction Techniques &
  1 &
  extraction \\
3 &
  Natural Product Bioactivity Evaluation &
  4 &
  compounds, DPPH, anti-inflammatory activity \\
3  & Phytochemical Extraction and Bioactivity Studies                 & 8  & antioxidant activity, methanol extract, ethanol extract, isolated compounds \\
3  & Phytochemical Antioxidants and Anti-inflammatory Agents Analysis & 12 & IC50 values, absolute configuration, spectroscopic analysis, NMR              \\
3 &
  Phytochemical Bioactivity Profiling &
  16 &
  flavonoids, spectrometry, anti-inflammatory, IC50 \\ \bottomrule
\end{tabular}%
}
\end{table}

\clearpage
\begin{landscape}
\renewcommand{\arraystretch}{1.5}
\begin{table}[]
\centering
\caption{
\textbf{Examples of labels generated by paper titles only.}
Curated examples sourced from two datasets, \textit{Oncology \& carcinogenics} (\#3211), and \textit{Artificial intelligence (\#4206)}.
The results of two prompts are shown.
The first details the labels created in the case that \textit{only prominent paper titles} are used to generate the labels. 
The second details the case that \textit{all cluster characteristics} are used.
Examples are selected among those clusters for which a label was rejected for being too vague. 
The rejected label and final label for the titles-only prompt are shown (``\textit{titles}''), alongside the final label for the all-characteristics (``\textit{all}''), and the concepts and paper titles used to create the labels. 
Concepts and paper titles are semicolon-delimited.
}
\label{si:tab:title_examples}
\small
\resizebox{\linewidth}{!}{%
\begin{tabular}{@{}|l|l|p{2.5cm}|p{2.5cm}|p{2.5cm}|p{12cm}|p{12cm}|@{}}
\textbf{Dataset} &
  \textbf{ID} &
  \textbf{Rejected (titles)} &
  \textbf{Label (titles)} &
  \textbf{Label (all)} &
  \textbf{Concepts} &
  \textbf{Titles} \\
3211 &
  8 &
  Epigenetic Cancer Therapy &
  Targeted Epigenetic Oncology &
  Onco-Cardiology &
  cardiomyopathy; heart failure;   myocarditis; cardiotoxicity; ejection fraction; arrhythmias; stroke;   cardiovascular events; antiplatelet therapy; acute coronary syndrome;   coronary syndrome; adverse cardiovascular events &
  Histone deacetylase inhibitors:   molecular mechanisms of action; Small-molecule MDM2 antagonists reveal   aberrant p53 signaling in cancer: Implications for therapy; A Phase 1 study   of RO6870810, a novel bromodomain and extra-terminal protein inhibitor, in patients   with NUT carcinoma, other solid tumours, or diffuse large B-cell lymphoma \\
3211 &
  55 &
  Hormonal and Molecular Pathology   of Breast Cancer &
  Breast Cancer Endocrine Therapy   and Genetic Risk Factors &
  Menopausal Oncology Therapies &
  menopausal symptoms; hormone   replacement therapy; replacement therapy; hormone therapy; Postmenopausal   Women; serotonin reuptake inhibitors; reuptake inhibitors; AI therapy;   menopause; postmenopausal women; arthralgia; hormone &
  Effects of Tamoxifen and   Exemestane on Cognitive Functioning of Postmenopausal Patients With Breast   Cancer: Results From the Neuropsychological Side Study of the Tamoxifen and   Exemestane Adjuvant Multinational Trial; Overexpressions of Cyclin B1, cdc2,   p16 and p53 in Human Breast Cancer: The Clinicopathologic Correlations and   Prognostic Implications; Is Hormone Replacement Therapy Safe in Women With a   BRCA Mutation? \\
3211 &
  64 &
  Cancer Genomics and   Transcriptomics &
  Oncogenic RNA and Cellular   Analysis &
  Fibroblast Dynamics in Cancer   Progression &
  cancer-associated fibroblasts;   fibroblasts; stem cells; tumor microenvironment; mice; tumor stroma;   stem-like cells; cancer stem cells; cells; cancer cells; cell lines;   properties of cancer cells &
  RNAscope A Novel in Situ RNA   Analysis Platform for Formalin-Fixed, Paraffin-Embedded Tissues; Mutant p53   Gain of Function in Two Mouse Models of Li-Fraumeni Syndrome; Single-Cell   Analysis Reveals Fibroblast Clusters Linked to Immunotherapy Resistance in Cancer \\
3211 &
  68 &
  Molecular Oncology of Cancer   Prognostics &
  Genomic Predictors of Cancer   Treatment Response &
  Molecular Genetics of Colorectal   Cancer &
  mismatch repair; genotypes;   polymorphism; colorectal cancer; p53 codon; colorectal cancer patients;   microsatellite instability; polymerase chain reaction-restriction fragment   length polymorphism; reaction-restriction fragment length polymorphism; codon   72; microsatellite; colorectal carcinoma &
  The Immune Biology of   Microsatellite-Unstable Cancer; Almost all articles on cancer prognostic   markers report statistically significant results; New advances in DPYD   genotype and risk of severe toxicity under capecitabine \\
3211 &
  71 &
  Cancer Molecular Therapeutics &
  Oncogenic Pathways and Molecular   Targets &
  Mitosis and DNA Damage Response   in Cancer Cells &
  replication stress; mitotic   spindle; checkpoint activation; mitosis; Chk1; mitotic arrest; checkpoint;   oxidative DNA damage; centrosome; DNA damage; cell lines; spindle &
  DNA double-strand break repair   pathway regulates PD-L1 expression in cancer cells; Small-molecule inhibitors   of human mitochondrial DNA transcription; Aurora-A Is Essential for the   Tumorigenic Capacity and Chemoresistance of Colorectal Cancer Stem Cells \\
4206 &
  6 &
  Neuro-Inspired Robotics and   Machine Learning &
  Spiking Neural Systems in   Robotics and Learning &
  Neuromorphic Computing and   Hardware Acceleration &
  computing architecture; hardware   accelerators; spiking neurons; deep learning compilers; energy efficiency;   GPU; GPGPU; FPGA; real-time processing; DSP; neural model; edge devices &
  A Survey of Robotics Control   Based on Learning-Inspired Spiking Neural Networks; Weighted Fuzzy Spiking   Neural P Systems; Scaling up machine learning \\
4206 &
  14 &
  Optimization in Quantum and   Cooperative Pathfinding Systems &
  Quantum and Cooperative   Multi-Agent Optimization &
  Multi-Agent Bandit Optimization   and Crowdsensing &
  regret; bandits; bandit problem;   regret bounds; multi-armed bandit problem; equilibrium; multi-armed bandit;   equilibrium strategies; mobile crowdsensing; crowdsensing; bandit algorithms;   sensing tasks &
  Interactively optimizing   information retrieval systems as a dueling bandits problem; Experimental   quantum speed-up in reinforcement learning agents; Finding Optimal Solutions   to Cooperative Pathfinding Problems \\
4206 &
  40 &
  Deep Learning Architectures and   Applications; Deep Learning Algorithms and AutoML &
  Review of Deep Learning and   AutoML Techniques &
  Intelligent Fault Diagnosis and   Detection Systems &
  fault diagnosis; fault; fault   tree; defect detection; diagnosis method; fuzzy Petri net; fault tree   analysis; Bayesian network; fault diagnosis method; diagnosis system;   automatic transfer lines; diagnosis strategy &
  Review of Deep Learning   Algorithms and Architectures; A review of deep learning with special emphasis   on architectures, applications and recent trends; AutoML: state of the art   with a focus on anomaly detection, challenges, and research directions \\
4206 &
  67 &
  Adaptive Control in Bio-Inspired   Robotics &
  Adaptive Bio-Robotics Control   Systems &
  Humanoid and Bipedal Robot   Locomotion &
  biped robot; humanoid robot;   humanoid; walking robot; quadruped robot; robot; walking; gait; pattern   generator; motion; ZMP; walking pattern &
  Fuzzy Approximation-Based   Adaptive Backstepping Control of an Exoskeleton for Human Upper Limbs; Fast,   Robust Quadruped Locomotion over Challenging Terrain; Online trajectory   generation in an amphibious snake robot using a lamprey-like central pattern   generator model \\
4206 &
  97 &
  Hybrid Evolutionary Algorithms &
  Hybrid Particle Swarm   Optimization &
  Particle Swarm Optimization   Algorithms &
  particle swarm optimization;   swarm optimization; benchmark functions; inertia weight; PSO algorithm;   particle swarm optimization algorithm; swarm; optimization particle swarm   optimization; swarm optimization algorithm; test functions; standard particle   swarm optimization; convergence &
  Orthogonal Learning Particle   Swarm Optimization; A hybrid genetic algorithm and particle swarm   optimization for multimodal functions; Genetic Learning Particle Swarm   Optimization
\end{tabular}%
}
\end{table}

\end{landscape}

\end{document}